\documentclass[a4paper, 12 pt] {article}

\usepackage{amsmath}
\usepackage{amsfonts}
\usepackage{amsthm}
\usepackage{amscd}

\usepackage{graphicx}
\usepackage{epsfig}
\usepackage{graphics}

\newcommand{\nvec}{{\bf n}}

\newcommand\grad{{\bf \nabla}}

\newcommand\evec{{\bf e}}
\newcommand\Nvec{{\bf N}}
\newcommand\zhat{{\bf  \hat z}}

\newcommand{\rvec}{{\bf r}}

\begin{document}

\title{Topology and Bistability in liquid crystal devices.}

% \affiliation command applies to all authors since the last
% \affiliation command. The \affiliation command should follow the
% other information
% \affiliation can be followed by \email, \homepage, \thanks as well.
\author{A Majumdar$^{\dag\ \ddag}$,
%\thanks{e-mail address: {\tt a.majumdar@bristol.ac.uk}} \
CJP Newton $^{\ddag}$,\ JM Robbins$^\dag$
%\thanks{e-mail address: {\tt j.robbins@bristol.ac.uk}} \
\& M Zyskin$^\dag$
\thanks{a.majumdar@bristol.ac.uk,chris.newton@hp.com,j.robbins@bristol.ac.uk, m.zyskin@bristol.ac.uk}\\
School of Mathematics\\ $^{\dag}$~ University of Bristol,
University Walk,
Bristol BS8 1TW, UK\\
and\\ $^{\ddag}$~
Hewlett-Packard Laboratories,\\
 Filton Road, Stoke Gifford, Bristol BS12 6QZ, UK}

\thispagestyle{empty}

\maketitle

\begin{abstract}
We study nematic liquid crystal configurations in a prototype
bistable device - the Post Aligned Bistable Nematic (PABN) cell \cite{kg}.
 Working within the Oseen-Frank continuum model, we describe the liquid crystal
configuration by a unit-vector field $\nvec$, in a model version
of the PABN cell. Firstly, we identify four distinct topologies in
this geometry. We explicitly construct trial configurations with
these topologies which are used as initial conditions for a
numerical solver, based on the finite-element method. The
morphologies and energetics of the corresponding numerical
solutions qualitatively agree with experimental observations and
suggest a topological mechanism for bistability in the PABN cell
geometry.

\end{abstract}

%\pacs{61.30.Jf,  11.10.Lm, 61.30.Dk, 61.30.Hn, 11.27.+d}

\newpage

\section{Introduction.}

Liquid crystals are an intermediate phase of matter between the
solid and liquid states. In the simplest liquid crystal phase, the
nematic phase, the constituent rod-like molecules tend to align
along a locally preferred direction. This mean direction of
molecular alignment is described by a director field
$\nvec(\rvec)$, which is an unoriented unit-vector field so that
the sign of $\nvec$ has no physical significance \cite{dg}.

The existence of a locally preferred direction and the resulting
anisotropic optical properties make liquid crystals very suitable
for display devices. Most of the liquid crystal displays (LCDs) in
use today, such as the Twisted Nematic and the Super Twisted
Nematic, are monostable \cite{sluckin}. They can support two
optically contrasting states, only one of which is stable without
an applied field. Recently, there has been considerable interest
in developing bistable display technologies, where there are two
or more stable, optically contrasting states
\cite{brown},\cite{kg}. Here, power is needed only to switch
between the different states but not to maintain them.

\begin{figure}[p]
%[hp]
\begin{center}
\includegraphics[width=3 in, height=3 in]{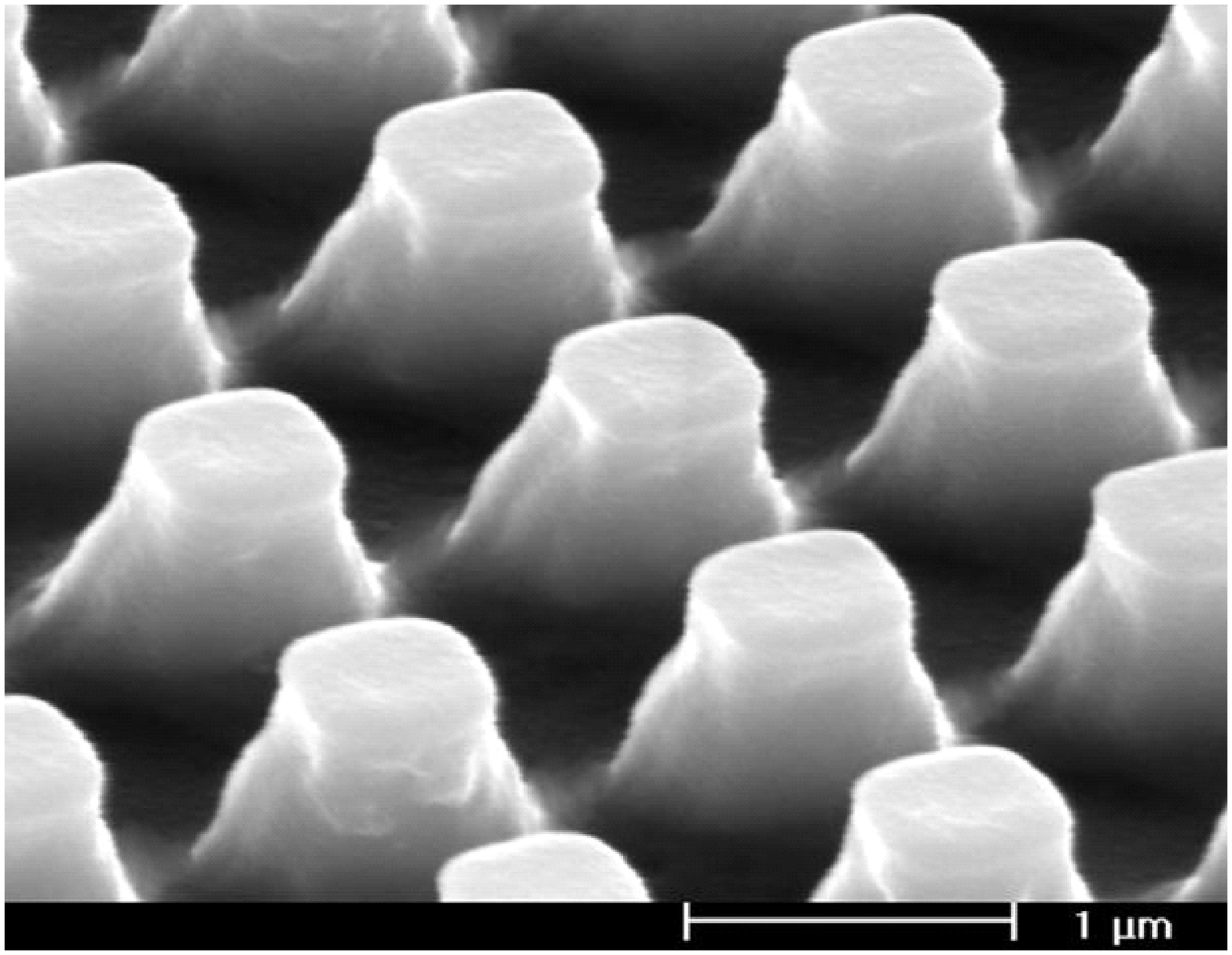}
\caption{The PABN device \cite{kg}.}\label{fig:PABN}
\end{center}
\end{figure}

Bistable nematic LCDs typically use a combination of complex
surface morphologies and surface treatments to stabilize the
different states \cite{brown,kg,davidson}. This paper focuses on
one such bistable device - the Post Aligned Bistable Nematic
(PABN) device \cite{kg}. The PABN device
consists of a liquid crystal layer sandwiched between two
substrates. The lower substrate is featured by an array of
microscopic posts, as shown in Figure~(\ref{fig:PABN}). The
boundary conditions are a mixture of tangent and normal conditions
in various parts of the geometry (referred to as
\emph{homogeneous} and \emph{homeotropic} respectively in liquid
crystal literature \cite{dg}). Tangent boundary conditions on a
surface constrain the director $\nvec$ to be in the plane of
the surface whereas normal boundary conditions constrain $\nvec$
to be perpendicular to the surface.

For a range of post heights, the PABN device is experimentally
observed to be bistable \cite{kg}. It supports two optically
contrasting states with long-term stability - one bright and the
other dark when viewed between crossed polarizers. Optical
modelling suggests that the dark state corresponds to a liquid
crystal configuration that tilts strongly around the microscopic
posts whereas the bright state corresponds to a suppressed tilt
profile around the posts ie. a comparatively planar configuration
\cite{kg}. The high-tilt configuration is referred to as the
\emph{tilted} state and the low-tilt configuration as the
\emph{planar} state in the remainder of the paper.

We use topological arguments to study static director
configurations in the PABN geometry. The admissible configurations
in this geometry can be partitioned into distinct topological
classes (for details see \cite{rz} and \cite{zyskin}) and one
might expect a stable configuration (or local energy minimizer)
for every such class. The key point is that topologically distinct
stable configurations cannot be continuously deformed into each
other and switching proceeds via creation or annihilation of
topological defects and boundary condition violation
\cite{yeomans, davidson}. Therefore, such topologically distinct
stable configurations, if they exist and have similar free
energies, lead to bistability in prototype devices.

Our study is related to basic questions about the existence and
properties of multiple equilibrium configurations in complex
geometries. For polyhedral geometries (such as a rectangular
prism), we have addressed several general questions about the
topological classification, energy estimates and regularity of
equilibrium configurations in previous work \cite{mrz2},
\cite{mrz4}. In this paper, we apply similar methods to a complex
non-convex polyhedral geometry - the PABN geometry.

The paper is organized as follows. In Section~(2), we describe our
simplified model of the PABN device, based on the Oseen-Frank
continuum model. In Section~(3), we identify four distinct (but
simple) topologies in this model geometry. We construct trial
configurations with these topologies in Section~(4). These trial
configurations are used as initial conditions for a numerical
solver of the equilibrium Euler-Lagrange equations (\ref{eq:EL}).
We obtain numerical solutions for each topological class. The
energies and morphologies of these solutions are then correlated
to the observed physical phenomena. The numerical results and
experimental observations are in qualitative agreement.

\section{Model.}

We work with a simplified model version of the PABN geometry,
neglecting factors such as post tilt and replacing the rounded
edges and corners in Figure~(\ref{fig:PABN}) by sharp features.
Further, we only consider configurations which are periodic in the
array so that it suffices to look at what happens around a single
post. (This is a realistic assumption since a typical pixel in the
PABN device consists of thousands of these posts.)

Our model geometry is displayed in Figure~(\ref{fig:modelgeom}) -
a single rectangular post of fixed square cross-section
($L_p\times L_p$) and variable height, $h$, inside a cell of fixed
dimensions, $L_c \times L_c \times H$. The post and cell
cross-sectional parameters, $L_p$ and $L_c$ respectively, are
chosen so that $L_c=2L_p$ whereas the cell height $H$ is fixed to
be $H=3L_c$. These choices roughly agree with the actual device
parameters.

We model the liquid crystal configuration, outside the rectangular
post, by a unit-vector field $\nvec(\rvec)$. We impose periodic
conditions on the cell boundaries so that
\begin{equation}
\nvec\left(x,y,z\right) = \nvec\left(x + L_c, y, z\right) =
\nvec\left(x, y + L_c, z \right) ~\textrm{etc.}
\label{eq:periodic}
\end{equation} The constraint
(\ref{eq:periodic}) allows us to extend our results to a periodic
array of posts. Further, $\nvec$ is taken to satisfy tangent
boundary conditions on the bottom substrate and post surfaces and
normal conditions on the top substrate, consistent with the
boundary conditions in the device.

Working within the Oseen-Frank continuum model, the liquid crystal
energy is given by
\begin{equation}
E\left[\nvec\right] = \int_V w\left(\nvec, \grad \nvec \right)~dV
\label{eq:Frank}
\end{equation} where
\begin{eqnarray}
&& w\left(\nvec, \grad \nvec \right) =
K_1\left(\grad\cdot\nvec\right)^2 + K_2\left(\nvec\cdot
\grad\times \nvec \right)^2 + K_3\left(\nvec \times \grad \times
\nvec \right)^2 \nonumber\\ && \qquad \qquad \qquad  + \left(K_2 +
K_4 \right)\left(\textrm{tr}(\grad\nvec)^2 -
\left(\grad\cdot\nvec\right)^2 \right) \label{eq:Frank2}
\end{eqnarray} and the $K_j$ are material-dependent elastic
constants \cite{dg}.  The stable configurations then correspond to
local minimizers of (\ref{eq:Frank}), subject to the imposed
boundary conditions.

\section{Topology.}

Our aim in this paper is to identify and analyse topologically
distinct stable configurations in the PABN cell (ie.
configurations that cannot be continuously deformed into each
other). In this section, we first study the two-dimensional
Zenithally Bistable Nematic (ZBN) cell as an illustrative example
\cite{brown, newtonspiller}. The ZBN cell has a comparatively
simple two-dimensional geometry where the topology is
characterized by a single quantity - the \emph{planar winding}
number \cite{techreport}. Here, we look at the experimentally
observed states and identify their distinct topologies. Then, we
study the three-dimensional PABN cell, where the geometry is far
more complex and the topology richer. We identify four different
topological classes in this geometry. These classes generate
tilted and comparatively planar profiles around the rectangular
post and, hence, serve as good candidates for the topologies of
the experimentally observed tilted and planar states.

\subsection{ZBN cell.}

The ZBN cell, like most liquid crystal cells, consists of a liquid
crystal layer sandwiched between two substrates. The bottom
substrate is planar whereas the upper substrate is a mono-grating.
Both substrates are treated to be homeotropic so that $\nvec$ is
constrained to be normal to these surfaces.

\begin{figure}[p]
\begin{center}
\includegraphics[width=3 in, height=3 in]{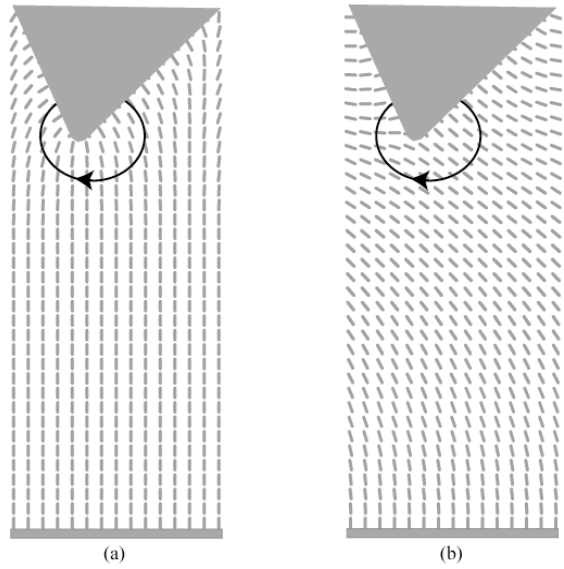}
\caption{(a) The high-tilt state.  (b) The low-tilt
state.}\label{fig:zbn}
\end{center}
\end{figure}

The ZBN cell supports two stable, optically contrasting
configurations - the high-tilt almost vertical state which is dark
when viewed under crossed polarizers and the low-tilt state,
supporting greater bulk distortion, that is bright under crossed
polarizers \cite{newtonspiller, brown} (tilt is measured with
respect to the horizontal direction). The cell cross-section (with
a wedge-shaped upper substrate) and the director profiles for the
two observed states are shown in Figures~(\ref{fig:zbn}a) and
(\ref{fig:zbn}b).

The topology of these states is characterized by the rotation of
$\nvec$ along a path connecting the two sides of the grating, as
shown in Figure~(\ref{fig:zbn}). The high-tilt state has the
minimum allowed rotation consistent with the boundary conditions
whereas the low-tilt state exhibits greater intermediate rotation.
In fact, the net rotations for the two states differ by $\pi$ or a
half-winding number \cite{techreport}. This then implies that the
two states have different winding numbers and are thus
topologically distinct, leading to their long-term stabilities.

It should be noted that $\nvec$ can, in principle, rotate by
arbitrarily large amounts around the relief grating (between the
fixed orientations at the sides) but such states have much higher
energies and are unlikely to be observed in practice.

\subsection{PABN cell.}

\begin{figure}[p]
\begin{center}
\scalebox{0.4}{\includegraphics{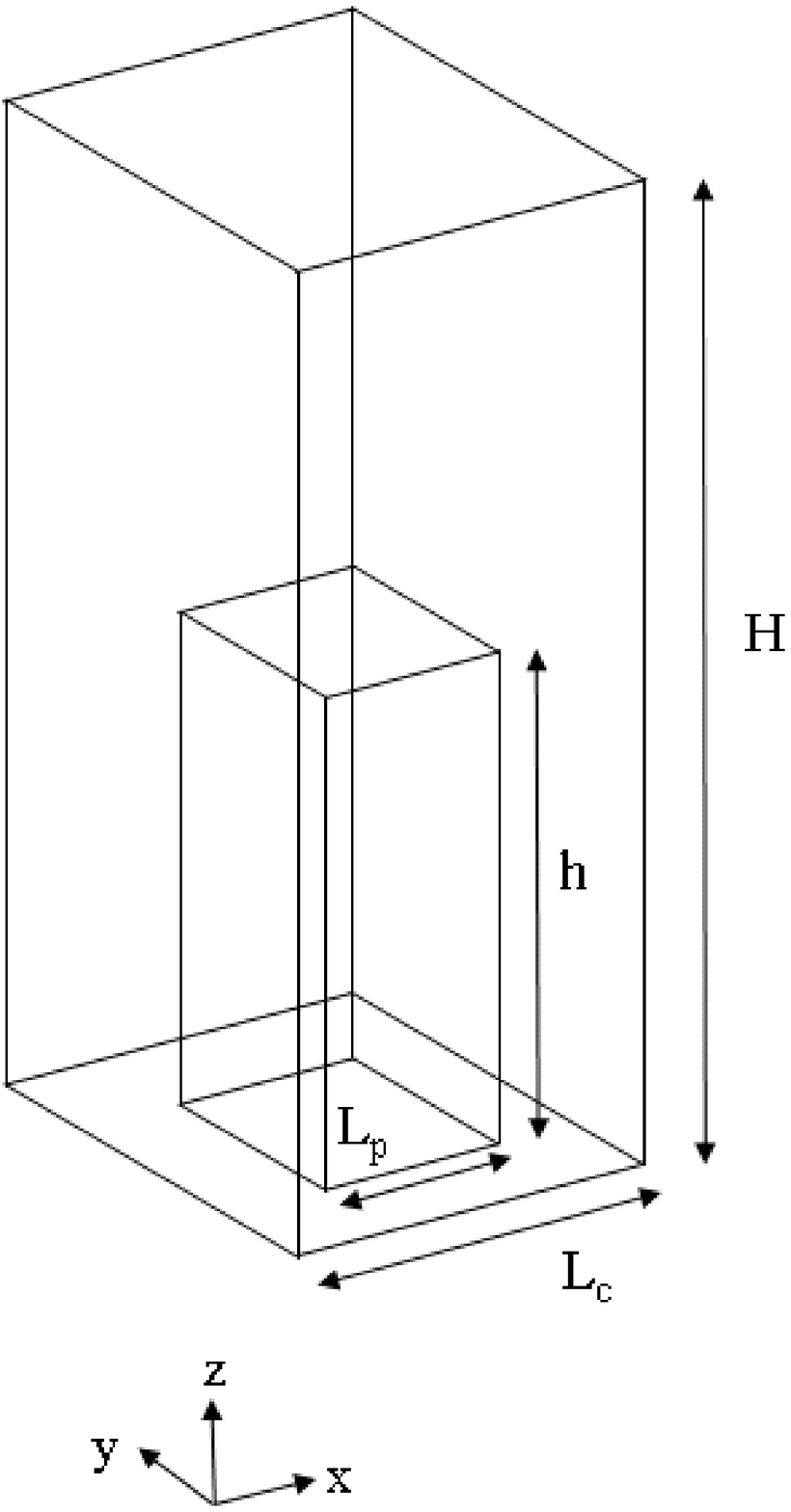}} \caption{The model
geometry with $L_c=1$, $L_p=0.5$ and $H=3$. When translated to
physical units, these correspond to $L_c = 1 \mu m$, $L_p = 500
nm$ and $H=3 \mu m$ respectively.}\label{fig:modelgeom}
\end{center}
\end{figure}

Next, we consider the three-dimensional model PABN cell, in
Figure~(\ref{fig:modelgeom}). Here, the topology cannot be merely
characterized by the planar winding number, as in the ZBN case. In
fact, there are at least three separate topological invariants on
the post edges, post faces and around the post vertices. In this
section, we do not give a detailed description of the topological
classification but identify four distinct, low-energy topologies.
A systematic account is given in \cite{rz} and \cite{zyskin}.

We first recall that the tangent boundary conditions on the bottom
substrate and the post surfaces imply that on these surfaces,
$\nvec$ takes values tangent to these surfaces. Therefore, on the
post edges, $\nvec$ is parallel to the edges and can take one of
two possible values. The value of $\nvec$, on a post edge, is
defined to be the corresponding edge orientation \cite{rz}.
$\nvec$ is necessarily discontinuous at the vertices, where three
or more edges meet. For simplicity, we only consider
configurations which are continuous everywhere away from the sharp
post vertices.

\begin{figure}[p]
\begin{center}
\scalebox{0.75}{\includegraphics{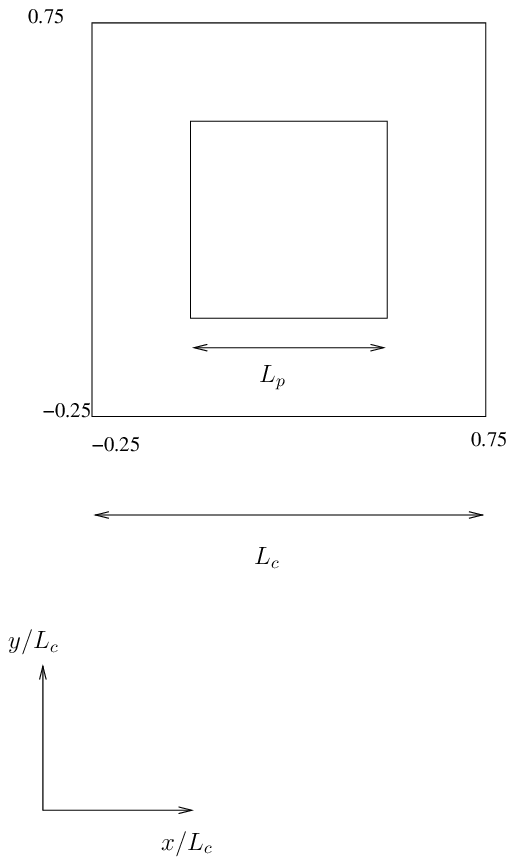}}
\caption{The post cross-section, $L_p \times L_p$, and the cell
cross-section, $L_c\times L_c$, where $L_c = 2L_p$.}
\label{fig:crosssection}
\end{center}
\end{figure}

\begin{figure}[p]
\begin{center}
\scalebox{0.6}{\includegraphics{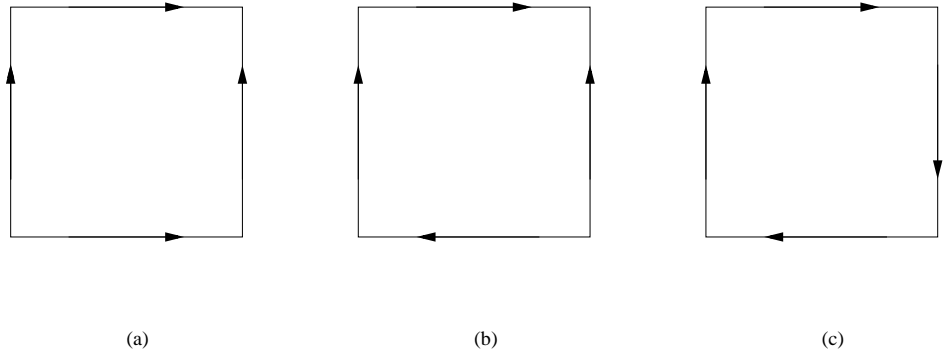}} \caption{(a) The most
symmetric choice. (b) Admissible choice of lower symmetry.(c)
Necessarily contains a planar defect.} \label{fig:xyedge}
\end{center}
\end{figure}

We first look at the top face of the post and the corresponding
four horizontal edges. Up to symmetry and the sign of $\nvec$,
there are three distinct choices of the horizontal edge
orientations on this face, as shown in Figure~(\ref{fig:xyedge}).
The last case necessarily creates a planar defect on this face and
is, therefore, excluded (the constraint of interior continuity
disallows certain choices of the edge orientations, for details
see \cite{rz}). Of the remaining two cases, we choose the first
one since it is more symmetric and is, therefore, expected to have
lower energy. Once the horizontal orientations on the top face are
fixed as in Figure~(\ref{fig:xyedge}a), the horizontal
orientations on the corresponding bottom edges are taken to
coincide with those on the top. This then completely determines
the edge orientations on the eight horizontal post edges in our
model, which are kept fixed for simplicity.

\begin{figure}[p]
\begin{center}
\scalebox{0.4}{\includegraphics{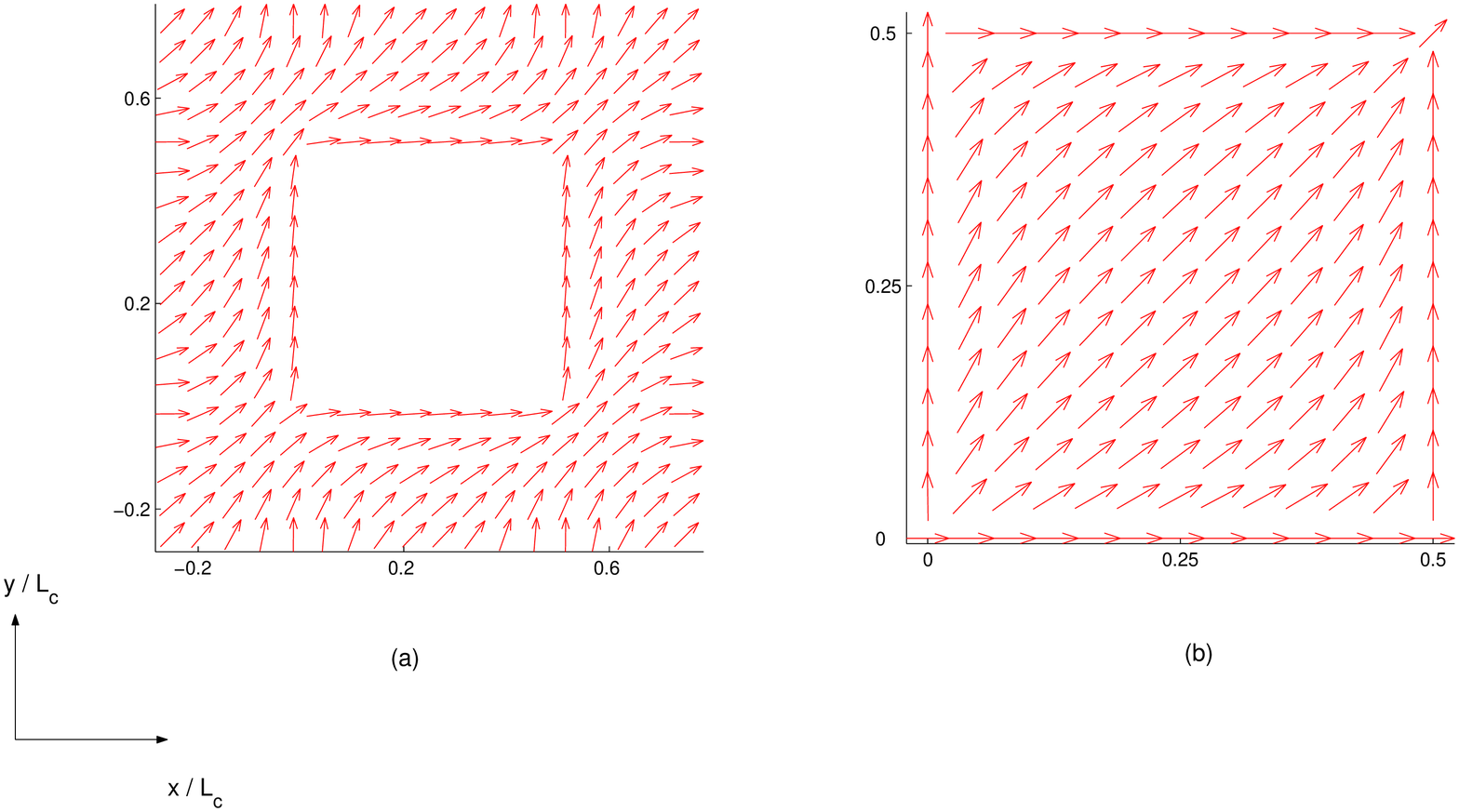}} \caption{(a) The
profile on the bottom substrate. (b) The profile on the top face
of the post.}\label{fig:postedge}
\end{center}
\end{figure}

\begin{figure}[p]
\begin{center}
\scalebox{0.4}{\includegraphics{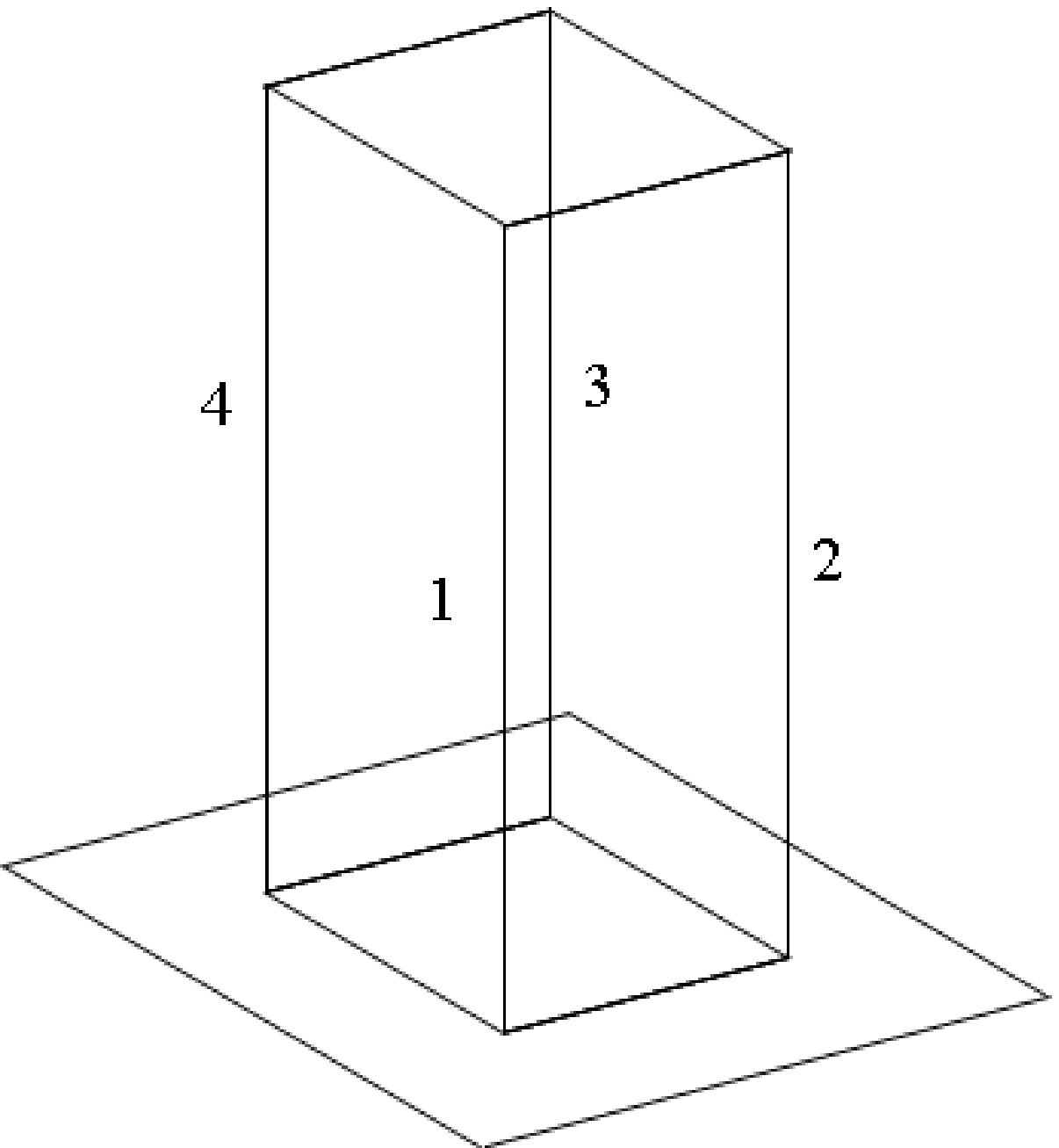}} \caption{ The four
vertical edges.}\label{fig:edge}
\end{center}
\end{figure}

We then consider the edge orientations on the four vertical post
edges. We recall that tangent boundary conditions imply that
$\nvec$ is either oriented upwards or downwards on these edges.
Labelling the four edges by $i=\left\{1, 2, 3, 4\right\}$ as in
Figure~(\ref{fig:edge}) and the corresponding edge orientation by
$\evec_i$, there are four distinct cases up to symmetry and the
sign of $\nvec$. These cases are enumerated in Table~(1). The
first case, $T$, corresponds to $\nvec$ being oriented upwards on
all four edges. For the second case, $P_1$, $\nvec$ is oriented
upwards on three vertical edges and downwards on the remaining
fourth edge whereas for the cases $P_2$ and $P_3$, $\nvec$ is
oriented upwards on two edges and downwards on the remaining two.
$P_2$ and $P_3$ are distinguished by the fact that $\evec_i$
changes sign on a pair of opposite faces in $P_2$ whereas
$\evec_i$ changes sign on all four vertical faces in $P_3$.

\begin{center}
\begin{table}
\begin{center}
\begin{tabular}{|c|c|c|c|c|}
  \hline
  % after \\: \hline or \cline{col1-col2} \cline{col3-col4} ...
   & $\evec_1$ & $\evec_2$ & $\evec_3$ & $\evec_4$ \\
  \hline
  $T$ & $\zhat$ & $\zhat$ & $\zhat$ & $\zhat$ \\
  \hline
  $P_1$ & $\zhat$ & $\zhat$ & $-\zhat$ & $\zhat$ \\
 \hline
  $P_2$ & $\zhat$ & $-\zhat$ & $-\zhat$ & $\zhat$ \\
  \hline
  $P_3$ & $\zhat$ & $-\zhat$ & $\zhat$ & $-\zhat$ \\
  \hline
  \end{tabular}
  \end{center}
  \caption{The four sets of vertical orientations.}
 \end{table}
 \end{center}

 In addition to the edge orientations, the topology of $\nvec$ is
 also determined by its behaviour on the post faces and around the
 post vertices. We take an arbitrary path connecting a pair of adjacent post edges, on a post face.
 The tangent boundary conditions imply that as we
 move along this path, $\nvec$ rotates in the plane of the corresponding
 face. We only consider topologies whereby $\nvec$ undergoes
 minimal rotation between pairs of adjacent post edges (which is
 $\pm\frac{\pi}{2}$ or a quarter-turn). In the terminology of
 \cite{rz}, we say that these topologies have zero \emph{kink numbers}.

The fixed horizontal edge orientations and the constraint of zero
kink numbers allows us to qualitatively predict the behaviour of
$\nvec$ on the bottom substrate and the top of the post, as shown
in Figure~(\ref{fig:postedge}a) and Figure~(\ref{fig:postedge}b).
On the bottom substrate, our choice of the horizontal orientations
implies that $\nvec$ splits at one of the vertices (the bottom
left vertex), follows the post edges and then rejoins at the top
right vertex, aligning along the square cross-sectional diagonal.
The net rotation between any two adjacent edges is just
$\pm\frac{\pi}{2}$, as is evident from
Figure~(\ref{fig:postedge}a). Similarly, on the top of the post
(refer to Figure~(\ref{fig:postedge}b)), we have a fractional
source defect at one of the vertices (the bottom left vertex),
accompanied by a sink defect at the diagonally opposite (top
right) vertex and any continuous $\nvec$ aligns along the
corresponding diagonal.

 The sharp post vertices can be treated as fractional
 point defects in our model; the corresponding strength is measured in terms of
 the neighbouring distortion around the vertex. We only consider
 topologies where $\nvec$ has minimal distortion
 around post vertices or equivalently, these vertices are fractional defects of minimal degree (the solid
 angle subtended by $\nvec$ as it varies around a post vertex is just $\pm\frac{\pi}{2}$).
 In the terminology of \cite{rz}, these
 topologies have minimal \emph{trapped areas}.

This then defines four distinct topologies, $\left\{T, P_1, P_2,
P_3 \right\}$, all of which have fixed horizontal orientations as
in Figure~(\ref{fig:xyedge}a), zero kink numbers and minimal
trapped areas but are distinguished by their vertical orientations
given in Table~(1). The topologies are labelled by their vertical
orientations for simplicity, since the vertical orientations are
the only topological parameters in our model.

\section{Numerical Modelling.}

We now investigate the existence of equilibrium stable
configurations with the four topologies cited in Section~(3). To
this end, we use the finite-element method to numerically solve
the Euler-Lagrange equations (\ref{eq:EL}) associated with the
Oseen-Frank energy (\ref{eq:Frank}) \cite{stewart}.The numerical
modelling is carried out in Femlab - a commercial partial
differential equation solver based on the finite-element method
\cite{Femlab}.

\begin{equation}
\frac{\partial}{\partial\nvec_i}w\left(\nvec, \nabla\nvec\right) =
\frac{\partial}{\partial
r_j}\left(\frac{\partial}{\partial\nvec_{i,j}}w\left(\nvec,
\nabla\nvec\right)\right) \label{eq:EL}
\end{equation} where $w\left(\nvec,\nabla\nvec\right)$ - the energy density has been
defined in (\ref{eq:Frank2}), $\nvec_i$ is the $i$-th component of
the unit-vector field and $\nvec_{i,j}$ is the $j$-th partial
derivative of $\nvec_i$.

We first construct trial configurations for each of the four
topologies. Here, it is simpler to specify an unnormalized vector
field $\Nvec=\left(N_x, N_y, N_z\right)$. The unit-vector field
$\nvec$ is then given by
\begin{equation}
\nvec = \frac{\Nvec}{\left|\Nvec \right|} .\label{eq:rep}
\end{equation}

The fixed horizontal orientations allow us to have a single
prescription for $N_x$ and $N_y$ in the four cases.
\begin{eqnarray}
&& N_x = \left(\sin\left(\frac{\pi x}{L_p}\right)\right)^2 \times
\left(\frac{H-z}{H}\right) \nonumber \\ && N_y
=\left(\sin\left(\frac{\pi y}{L_p}\right)\right)^2 \times
\left(\frac{H-z}{H}\right) \label{eq:IC1}
\end{eqnarray} where $L_p$ is the post cross-sectional parameter
and $H$ is the cell height.

For $0\leq z \leq h$, up to the height of the post, $N_z$ clearly
has different forms to account for the different sets of vertical
orientations.
\begin{equation}
  \label{eq:IC2}
  N_z =
  \begin{cases}
    z\times (h - z), & T,\\
%    s(w,\wbar)
 \left(z(h-z)\right)\times\left(1 + \cos\left(\frac{\pi
x}{L_p}\right) + \cos\left(\frac{\pi y}{L_p}\right)\right),&
    P_1,\\
    \left(z(h-z)\right )\times\cos\left(\frac{\pi
x}{L_p}\right),& P_2,\\ \left(z (h - z) \right)\times
\cos\left(\frac{\pi x}{L_p}\right)\times \cos\left(\frac{\pi
y}{L_p}\right)& P_3.
  \end{cases}
\end{equation} For $h \leq z \leq H$ (from the top of the post to
the top substrate), $N_z$ is given by
\begin{equation}
N_z = \frac{z - h}{H - h}  \label{eq:IC3}
\end{equation} for all topologies. (Note that for $L_c = 2L_p$ and our domain - $\frac{-1}{4}\leq \frac{x}{L_c}, \frac{y}{L_c} \leq \frac{3}{4}, 0\leq z\leq H$, the only points where $\Nvec$ vanishes are given by the post vertices,
which are necessarily singularities because of the tangent
boundary conditions.)

Given equations (\ref{eq:IC1}), (\ref{eq:IC2}) and (\ref{eq:IC3}),
one can easily check that the representative (\ref{eq:rep})
satisfies the boundary conditions with the correct topology and is
continuous away from the sharp post vertices. We use these trial
configurations as initial conditions for the numerical solver. We
numerically solve the Euler-Lagrange equations on a variable mesh
that allows for greater resolution near the sharp post features.
The numerical solver respects the boundary conditions and topology
so that the numerical solutions have the same topology as the
initial condition.

\begin{figure}[p]
\begin{center}
\scalebox{0.65}{\includegraphics{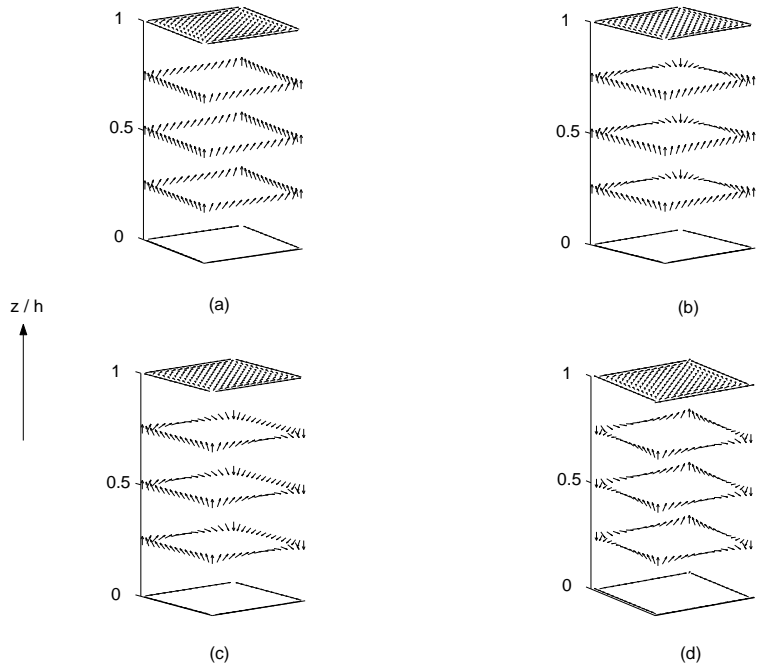}} \caption{(a) The $T$
profile. (b) The $P_1$ profile. (c) The $P_2$ profile. (d) The
$P_3$ profile.}\label{fig:postprofile}
\end{center}
\end{figure}

\begin{figure}[p]
\begin{center}
\scalebox{0.6}{\includegraphics{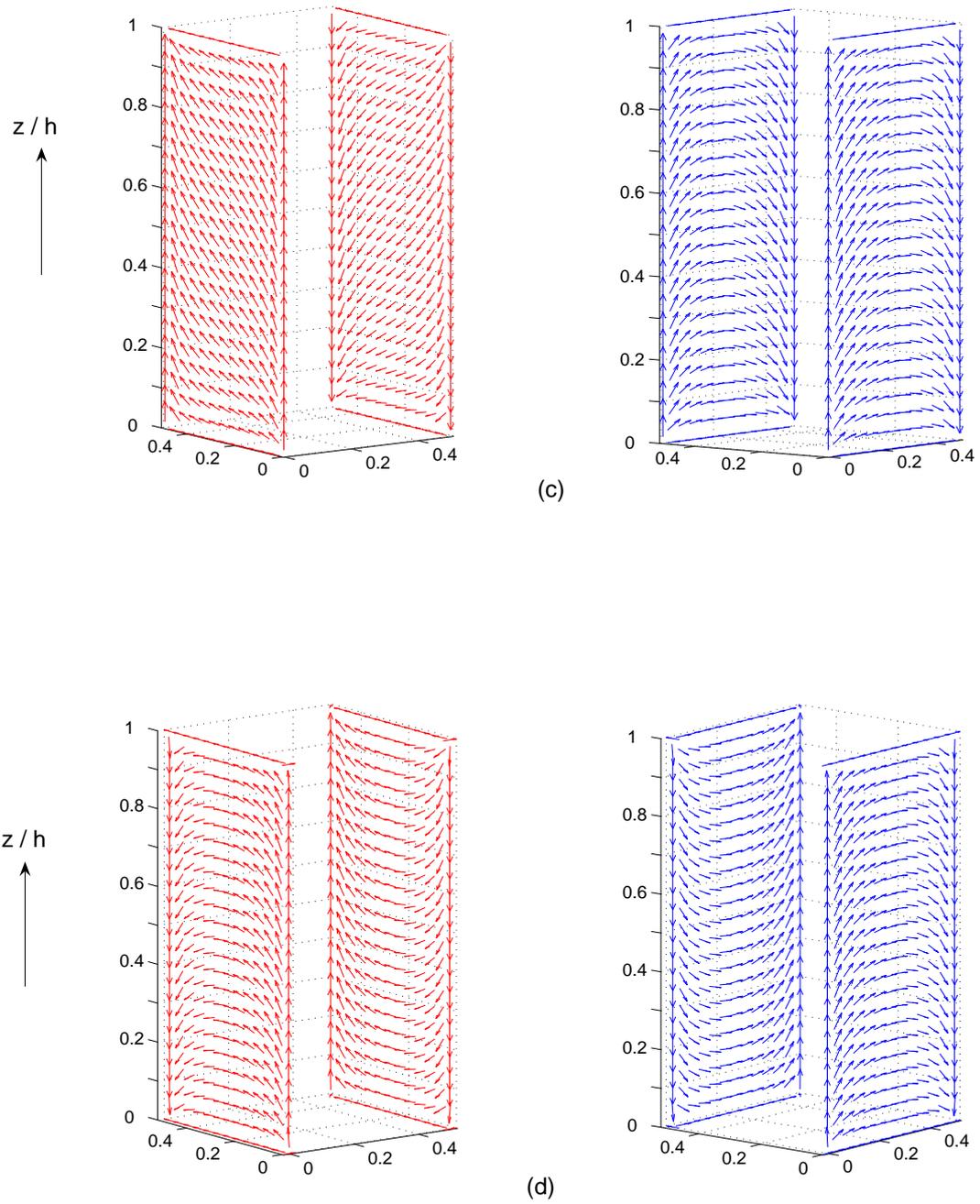}}
\end{center}\caption{The solution profiles on the faces. (a) $T$ (b) $P_1$ (c) $P_2$ (d) $P_3$}\label{fig:faceprofile}
\end{figure}

\begin{figure}[p]
\begin{center}
\scalebox{0.45}{\includegraphics{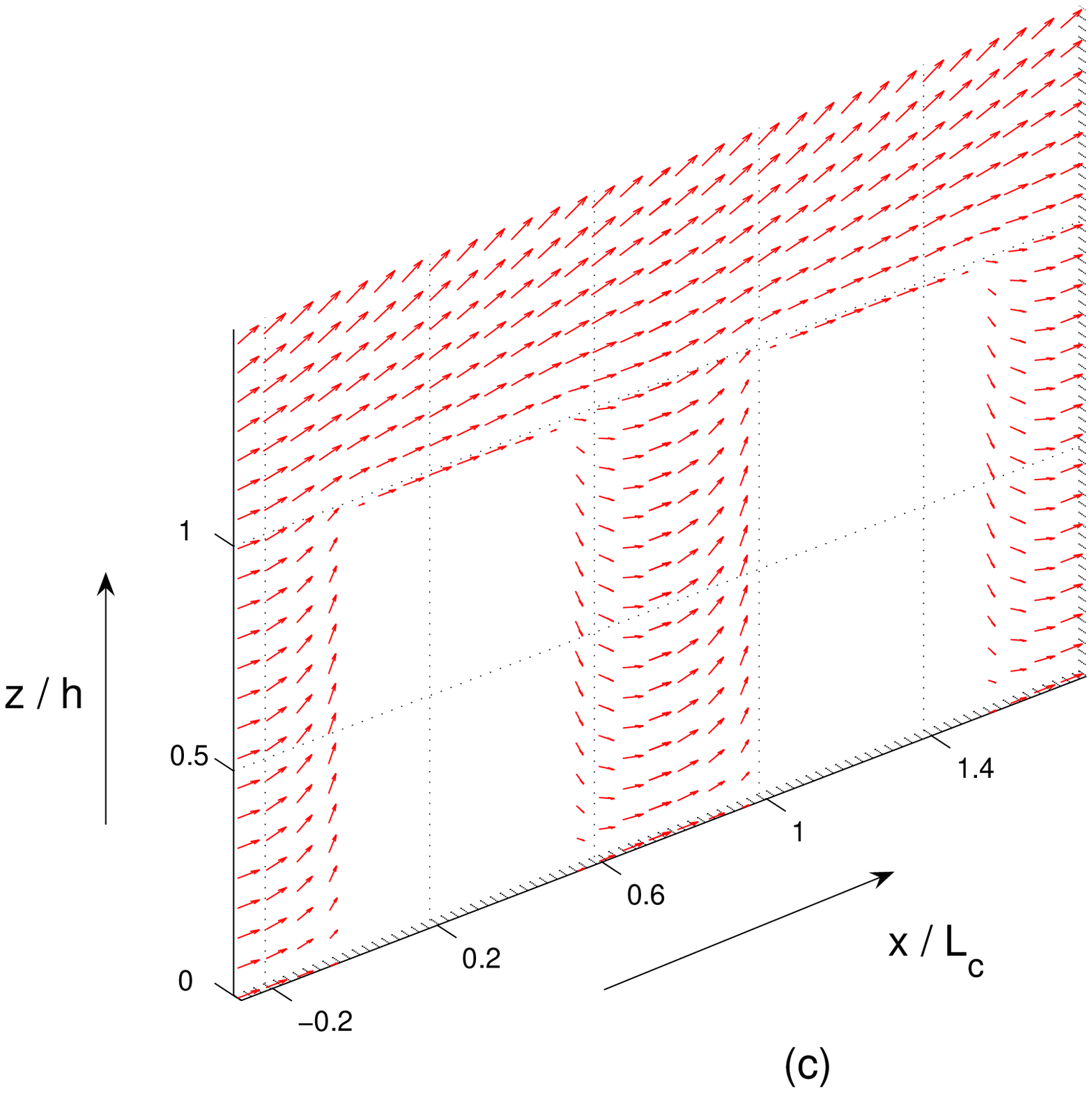}}
\scalebox{0.45}{\includegraphics{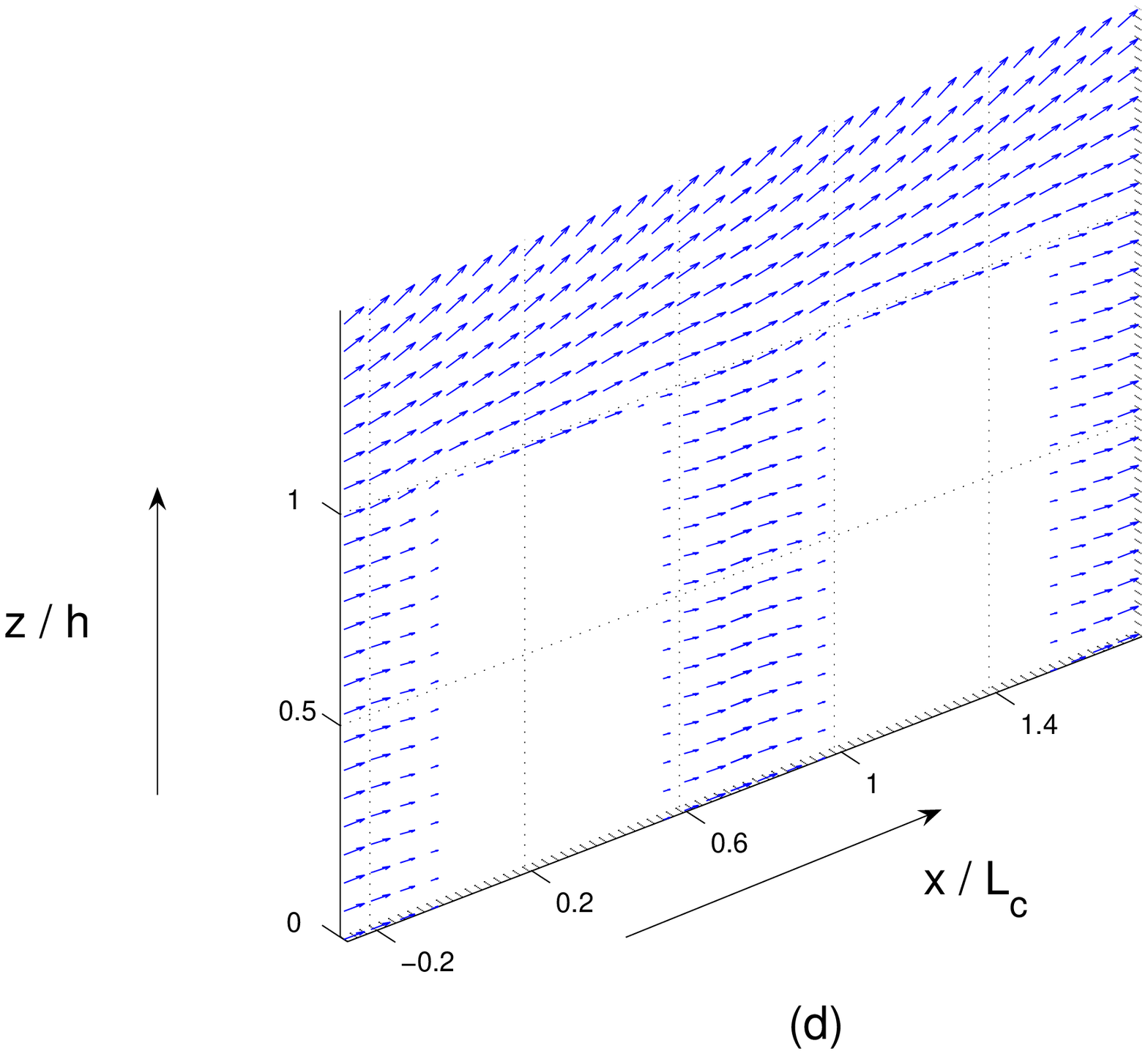}}
\caption{(a) The $T$ profile. (b) The $P_1$
profile. (c) The $P_2$ profile. (d) The $P_3$ profile. These plots
are the projection of the solution profiles onto the plane
$y=L_c/4$ i.e. only the $x$ and $z$ components of the profiles are
shown here. The dots, in particular, correspond to points where
the solution is oriented in the $y$-direction. Note that we only
plot up to height $z=3h/2$ because the profiles have similar
structure in the remainder of the cell $\frac{3h}{2}\leq z \leq
H$, where $h$ is the post height and $H$ is the cell height in
Figure~(\ref{fig:modelgeom}).}\label{fig:vectorprofile}
\end{center}
\end{figure}

We show the solution profiles with $h=L_c$ in
Figures~(\ref{fig:postprofile}), (\ref{fig:faceprofile}) and
(\ref{fig:vectorprofile}). We plot the profile around the
rectangular post for $\left\{T,P_1,P_2,P_3\right\}$ in
Figures~(\ref{fig:postprofile}a - d). It is evident that the
profiles are continuous and regular around the post with no point
or line singularities. In particular, there are no defects along
the leading and trailing edges, as suggested by previous modelling
\cite{kg}. In Figures (\ref{fig:faceprofile}a - d), we describe
the four profiles on the post faces in greater detail.
Figure~(\ref{fig:faceprofile}a) clearly shows that the $T$ profile
tilts upwards at all points as we move across the post faces and
aligns along the post diagonal. Figures~(\ref{fig:faceprofile}b-d)
correspond to the $P_1$, $P_2$ and $P_3$ profiles respectively.
Here, we note that whenever the vertical orientation changes sign
between a pair of successive vertical edges on a post face, there
is necessarily an intermediate planar region. This planar region
is typically a curve across the face, between the two edges in
question. We can clearly see these planar regions on a pair of
adjacent faces in the $P_1$ profile, on a pair of opposite faces
in the $P_2$ profile and on all four vertical faces in the $P_3$
profile.

\begin{figure}[p]
\begin{center}
\scalebox{0.5}{\includegraphics{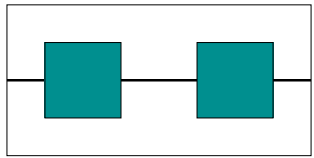}} \caption{A
periodically extended domain containing two posts. This is the
bottom cross-section, with the filled regions corresponding to the
post bases. The bold line indicates the cross-section $y=L_c/4$,
which is plotted in
Figure~(\ref{fig:vectorprofile}).}\label{fig:inset}
\end{center}
\end{figure}

In Figure~(\ref{fig:vectorprofile}), we plot the solutions on a
periodically extended domain, comprising of a pair of posts, along
the cross-section $y=L_c/4$ as shown in Figure~(\ref{fig:inset}).
The periodic boundary conditions imply that the profiles are
simply repeated between neighbouring posts. In particular, the
solution topology is the same for every post. This is illustrated
in Figure~(\ref{fig:vectorprofile}a), where the $T$ profile tilts
upwards between a pair of neighbouring posts, whilst interpolating
between the fixed orientations on the successive vertical edges.
Similarly, the planar regions observed in
Figures~(\ref{fig:faceprofile}b-d) extend between neighbouring
posts in Figures~(\ref{fig:vectorprofile}b-d), by virtue of the
periodic boundary conditions. Above the post, for $h\leq z \leq
H$, all profiles tilt upwards and finally become homeotropic on
the top substrate. These profiles, though discussed for a fixed
post height $h=L_c$, are characteristic of the solution profiles
for all $h$.
\begin{figure}[p]
\begin{center}
\scalebox{0.6}{\includegraphics{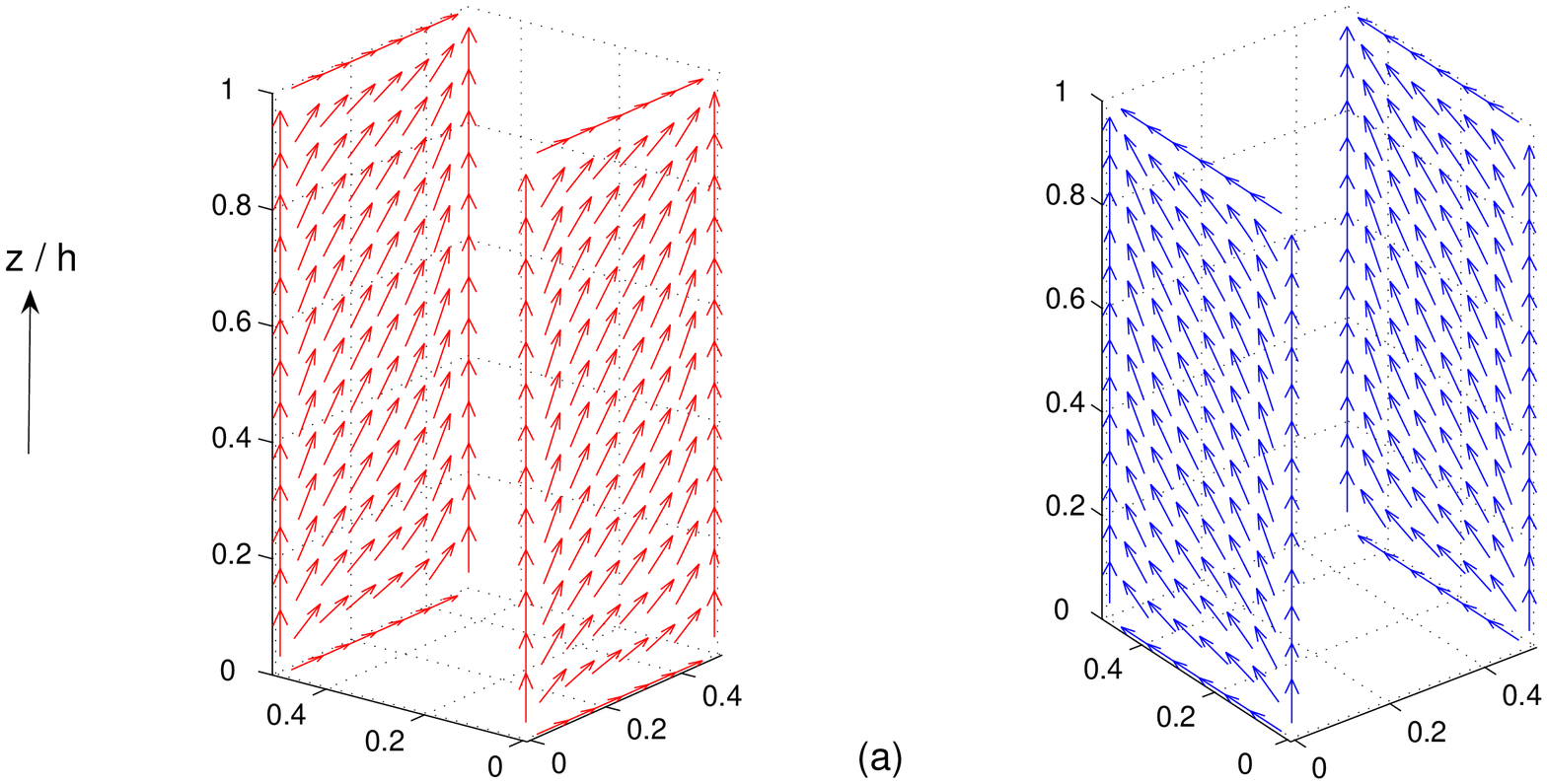}}
\end{center}
\end{figure}
\begin{figure}[p]
\begin{center}
\scalebox{0.6}{\includegraphics{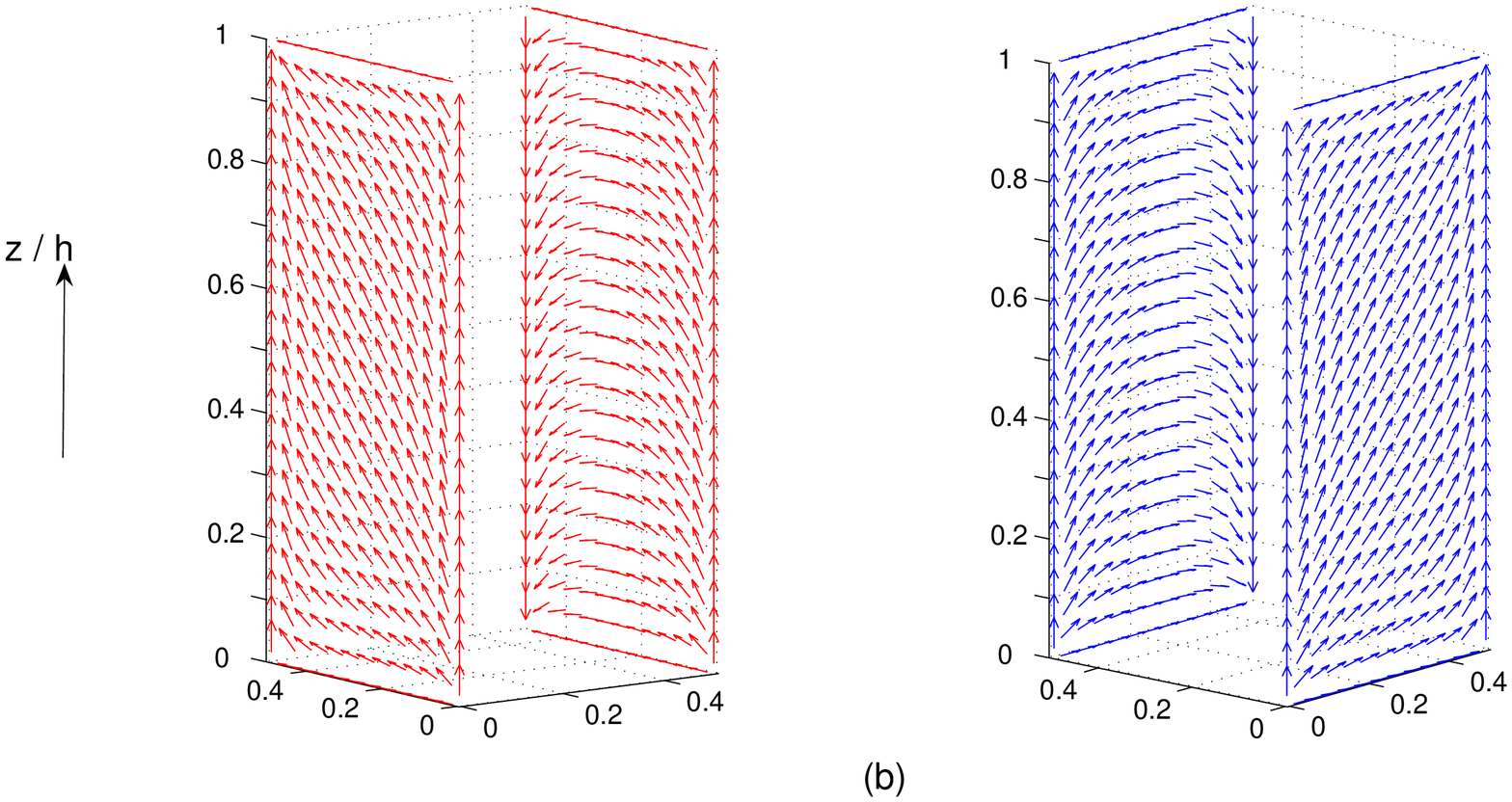}}
\end{center}
\end{figure}
\begin{figure}[p]
\begin{center}
\scalebox{0.6}{\includegraphics{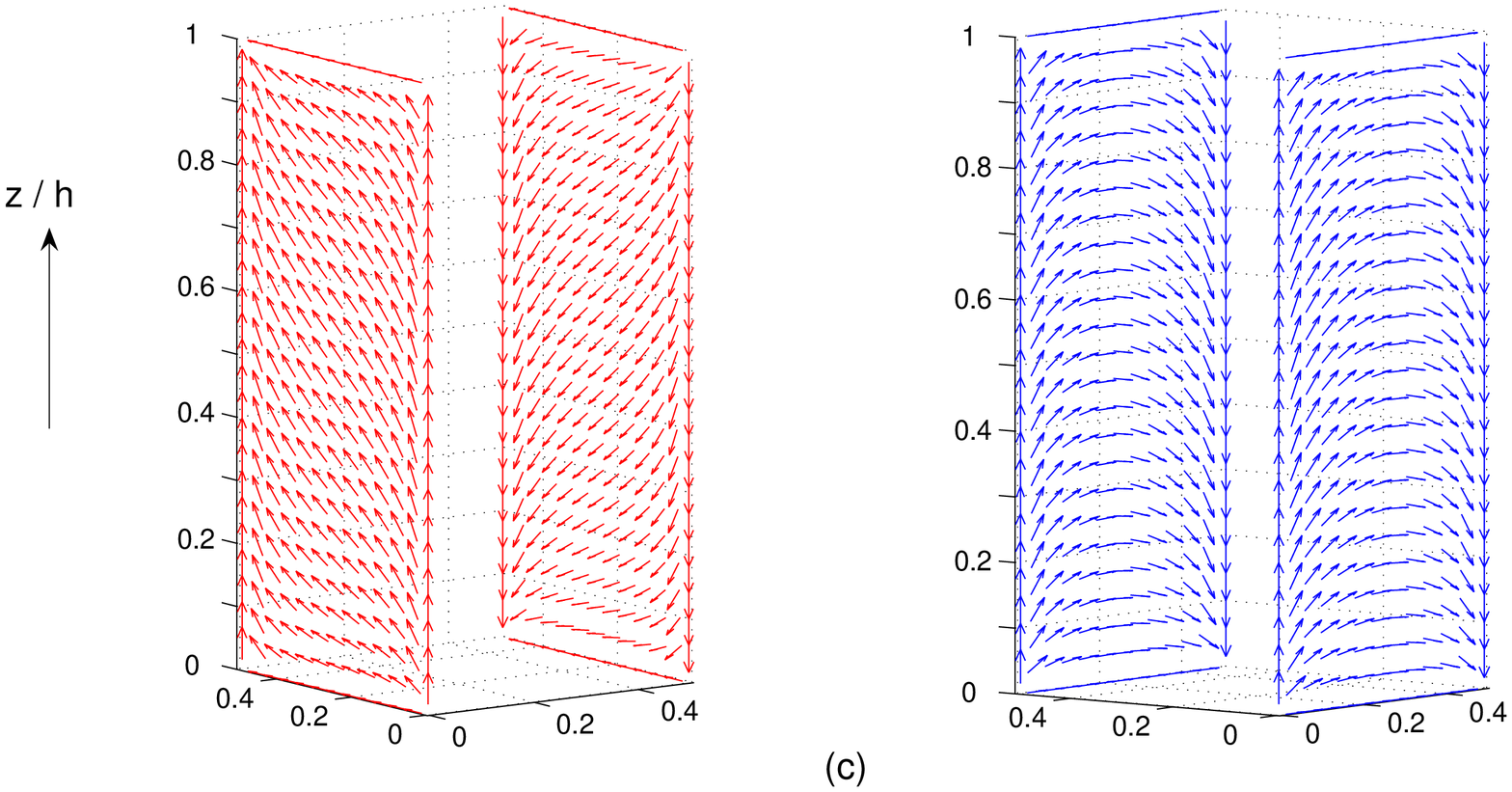}}
\end{center}
\end{figure}

\begin{figure}[p]
\begin{center}
\scalebox{0.5}{\includegraphics{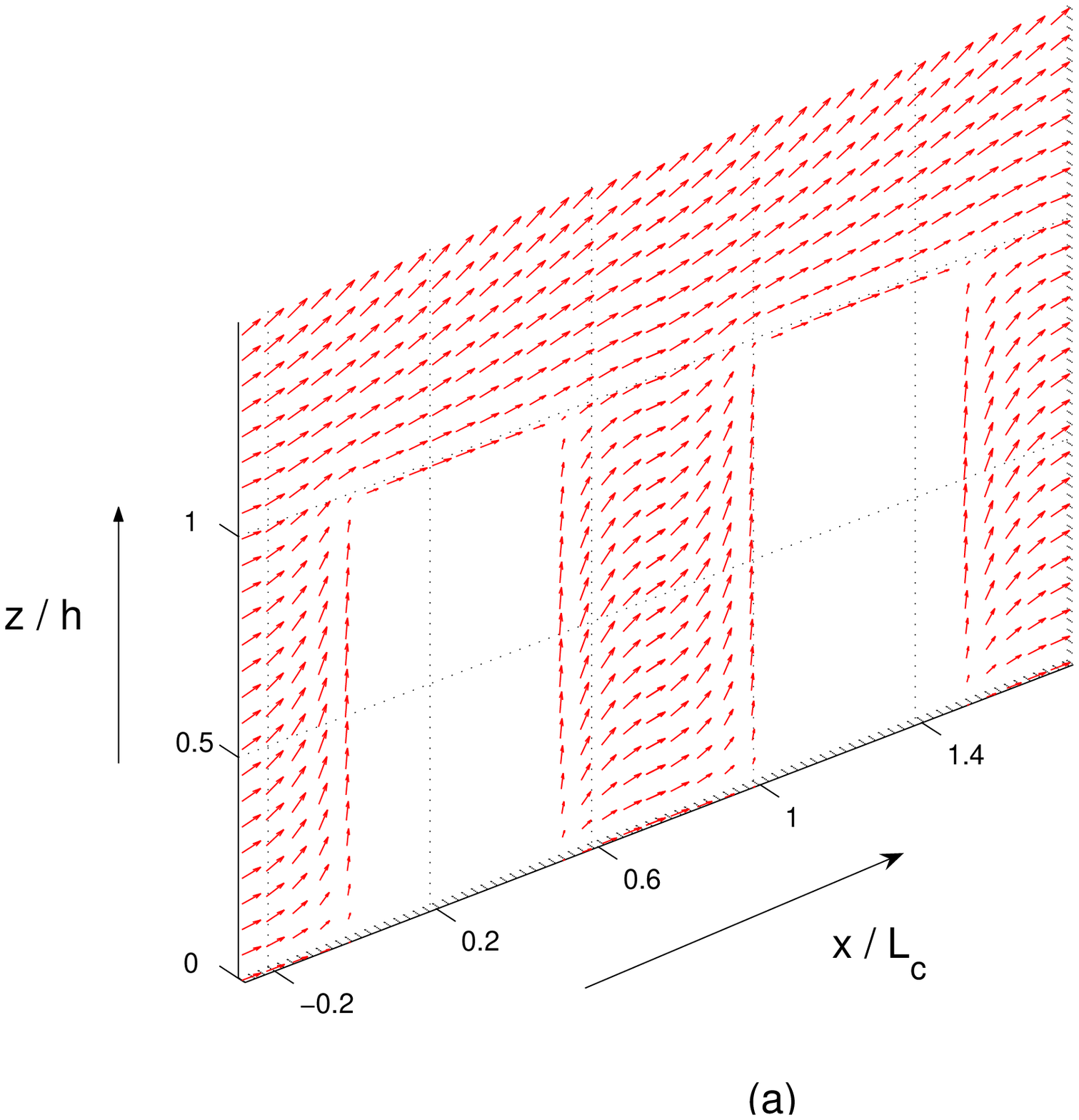}}
\scalebox{0.4}{\includegraphics{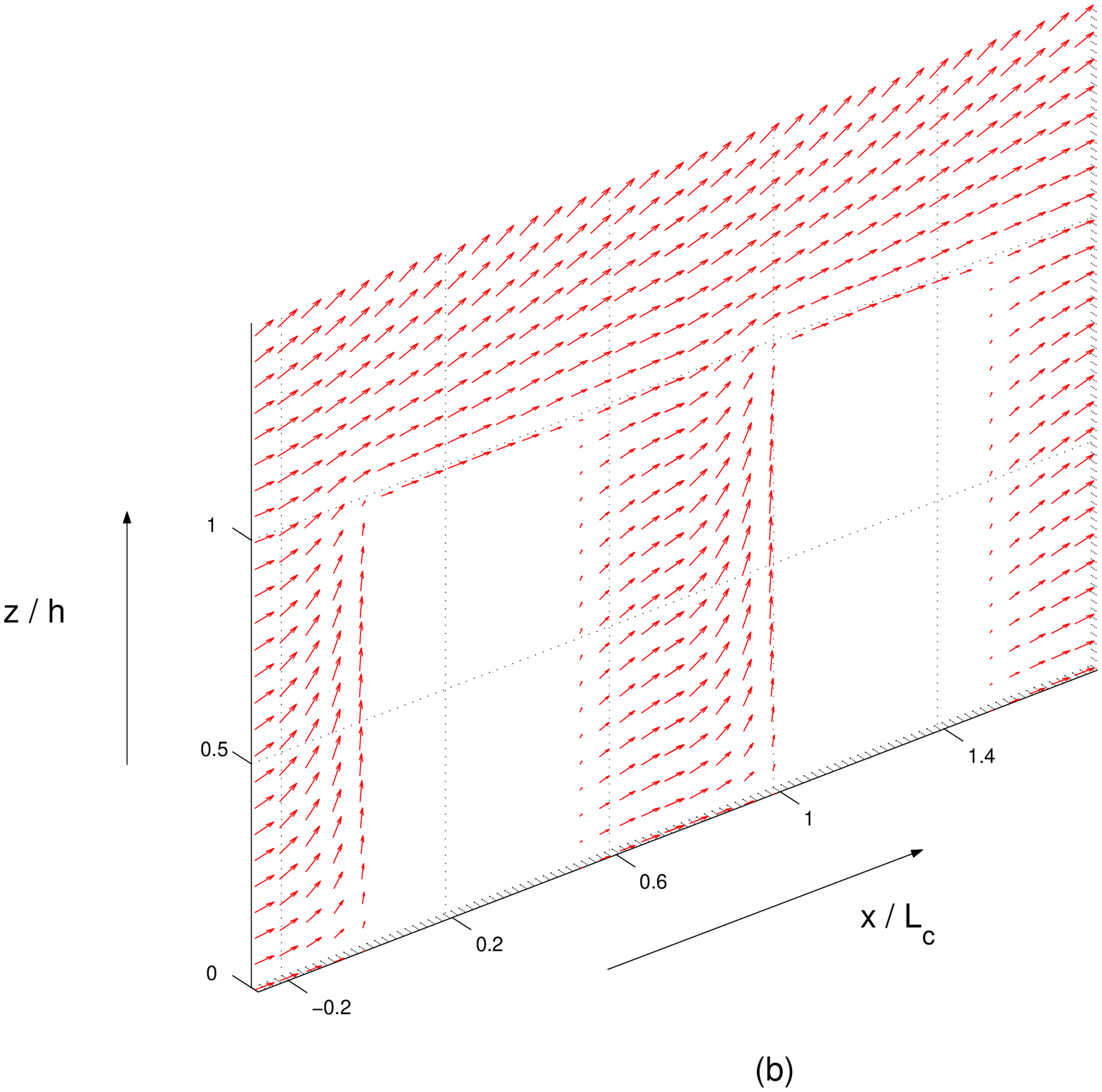}}
\end{center}
\end{figure}

\begin{figure}[p]
\begin{center}
\scalebox{0.4}{\includegraphics{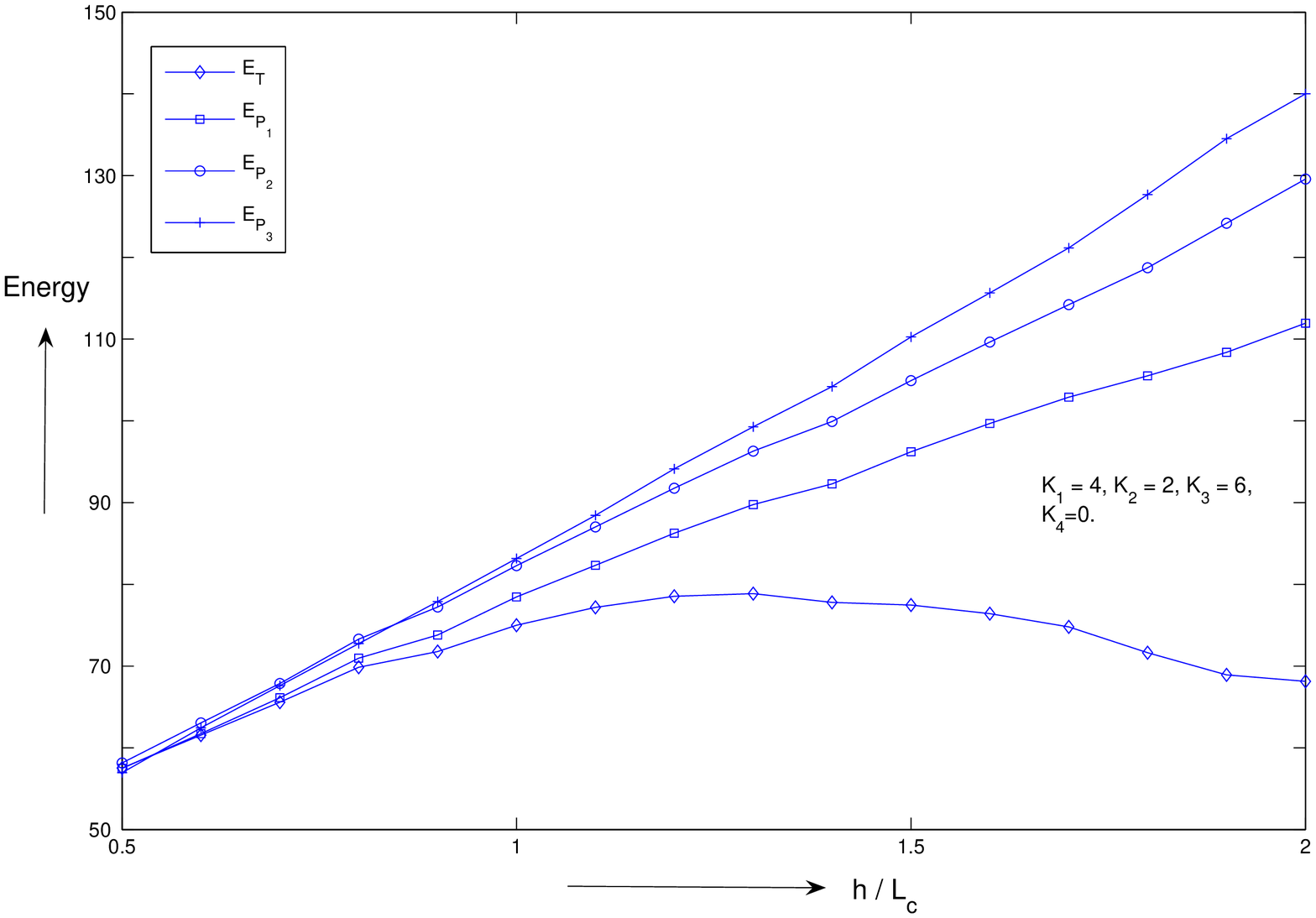}} \caption{The
solution energies in the Oseen-Frank case with $K_1=4$, $K_2=2$
and $K_3=6$.}\label{fig:3C}
\end{center}
\end{figure}

\begin{figure}[p]
\begin{center}
\scalebox{0.4}{\includegraphics {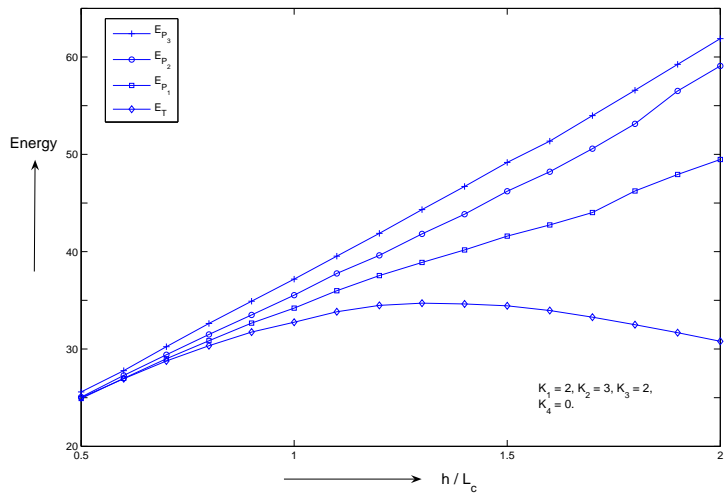}} \caption{The
solution energies in the one-constant approximation with $K_1 =
K_2 = K_3 = 2$.}\label{fig:OC}
\end{center}
\end{figure}

Next we look at how the solution energy varies with scaled post
height, $h/L_c$, for two different sets of elastic constants in
Figures~(\ref{fig:3C}) and (\ref{fig:OC}). The second set, with
equal elastic constants, corresponds to the widely used
\emph{one-constant} approximation \cite{dg,stewart}. It is evident
from Figures~(\ref{fig:3C}) and (\ref{fig:OC}) that the
qualitative trends for the two sets of elastic constants are the
same. However, the one-constant approximation in
Figure~(\ref{fig:OC}) is found to be computationally less
demanding than the unequal constant case and allows for greater
numerical resolution.

\begin{figure}[p]
\begin{center}
\scalebox{0.4}{\includegraphics {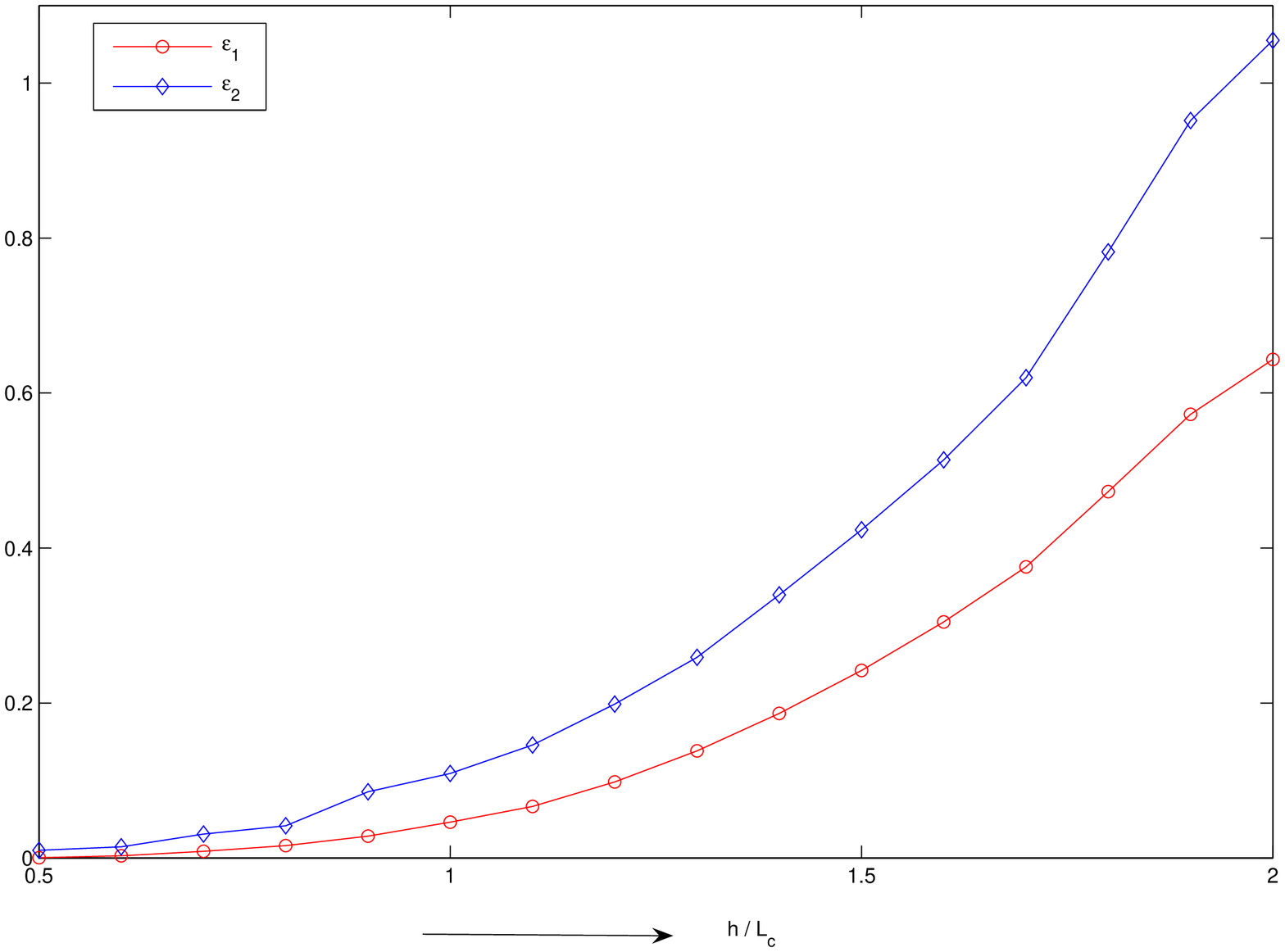}}\caption{Computing
$\epsilon_1$ and $\epsilon_3$ for the unequal constant case in
Figure~(\ref{fig:3C}).} \label{fig:deltaE}
\end{center}
\end{figure}

We first note that the $T$ solutions have the smallest energy
whereas the $P_3$ solutions have the highest (there is a slight
crossover between the $P_2$ and $P_3$ solutions in
Figure~(\ref{fig:3C}) but the free energy difference is
negligible). This is consistent with experimental observations
which show that the liquid crystal always relaxes into the
high-tilt state, when cooled down from the isotropic state
\cite{kg}.

These energy trends can be anticipated and explained on
topological grounds. As is evident from
Figures~(\ref{fig:faceprofile}a) and (\ref{fig:vectorprofile}a),
the $T$ solution has minimal distortion consistent with the
boundary conditions. The $P_3$ solution, on the other hand, has
the most strongly distorted profile because of the planar regions
on every vertical post face and between every pair of neighbouring
posts. These planar regions arise from the continuous
interpolation between opposite vertical orientations; these
vertical orientations, in turn, are part of the topological
characterization of these solution profiles. The minimal and
maximal distortion, in the $T$ and $P_3$ solutions respectively,
then qualitatively explain the energy trends.

We look at the free energy differences in
Figure~(\ref{fig:deltaE}). Here, we plot two quantities
\begin{equation}
\epsilon_1 = \frac{E_{P_1} - E_T}{E_{T}} \label{eq:delta1}
\end{equation} and
\begin{equation}
\epsilon_3 = \frac{E_{P_3} - E_{T}}{E_{T}} \label{eq:delta2}
\end{equation} for the elastic constants in
Figure~(\ref{fig:3C}).  $E_{P_1}$ is typically the smallest of
$\left\{E_{P_1}, E_{P_2}, E_{P_3}\right\}$ and $E_{P_3}$ is the
largest. It is obvious from (\ref{fig:deltaE}) that both $\epsilon_1$
and $\epsilon_3$ can be qualitatively described as piecewise linear
functions of $\frac{h}{L_c}$, with the gradient changing as a
function of post height. The gradient is smallest
in the interval
$\frac{h}{L_c}\in\left[0.5,0.9\right]$, where both $\epsilon_1$ and
$\epsilon_3$ are small and grow very slowly. We refer to this
interval, $h\in\left[0.5\times L_c, 0.9\times L_c\right]$, as the
\emph{plateau} region in the remaining discussion. For post
heights $h\geq L_c$, the gradient of both curves increases
appreciably so that $\epsilon_1$ and $\epsilon_3$ grow relatively quickly
with post height. This change in behaviour as a function of post
height can be explained by Figure~(\ref{fig:3C}). $E_T$, the
energy of the tilted solution, typically reaches a shallow local
maximum and then saturates as the post height increases. On the
other hand, both $E_{P_1}$ and $E_{P_3}$ grow linearly with post
height. These differences in the energy trends manifest in
$\epsilon_1$ and $\epsilon_3$, given by (\ref{eq:delta1}) and
(\ref{eq:delta2}), leading to the features cited above.

To understand what these numerical calculations mean in the
context of the actual device, we first note that the tilted and
planar states are observed experimentally for post heights in the
range $0.6$ microns to $1.2$ microns [private communication]. This
bistable region corresponds to $h\in\left[0.6\times L_c, 1.2\times
L_c\right]$ in our model. Further, as the posts become taller, the
lower energy tilted state is more easily observed than the planar
state and it becomes increasing difficult to switch from the
tilted state to the planar state \cite{kg}.  Our modelling shows a
plateau region for $\frac{h}{L_c}\in\left[0.5,0.9\right]$, where
both $\epsilon_1$ and $\epsilon_3$ are small and do not change
significantly. This plateau region suggests that we might observe
the topologically distinct $T$ and $P_i$ states, $i=1,2,3$, for
$\frac{h}{L_c}\in\left[0.5,0.9\right]$. Our numerical range and
the experimentally found bistable region are in qualitative
agreement. Secondly, the fact that $\epsilon_1$ and $\epsilon_3$ grow
quickly as a function of $\frac{h}{L_c}$ indicates that the $P_i$
states become energetically far more expensive than the $T$ states
as the post height increases. This, in turn, qualitatively
explains the experimental observations and switching
characteristics for taller posts. A more detailed analysis of the
liquid crystal energy would include surface effects and scalar
order parameters \cite{dg, stewart}.

\section{Conclusion.}
The methods and results presented in this paper illustrate how
topology can be exploited in order to find different static liquid
crystal configurations in prototype device geometries. For the
specific case of the PABN geometry, we have identified four
simple, low-energy topologies. This topological information is
then used to construct suitable initial conditions for a
finite-element numerical algorithm, yielding four distinct classes
of numerical solutions. It would be difficult to numerically find
these different solutions without the topological insight. Our
modelling suggests that the tilted and planar states in the PABN
geometry are topologically distinct, leading to a topological
mechanism for the observed bistability. (For example, the $T$
topology defined in Section~(3) is a good candidate for the
topology of the tilted state whereas the $P_i$ topologies serve as
good candidates for the planar state.) Secondly, these
 numerical solutions have comparable free energies over a certain range of post
heights, indicating a limited bistability region that is again
commensurate with experimental observations.

The methods outlined in this paper can be extended to more general
polyhedral geometries. We can use topology to organize the space
of admissible configurations, understand the structure of the
stable configurations and their energetics. This will be
investigated further in subsequent work.

\emph{Acknowledgements.} The preliminary Q-tensor modelling
results of CJN and Dr.Nigel Mottram suggested a possible topological
bistability mechanism in the PABN device, which we have
numerically investigated and quantified in this paper. We thank
them for sharing their modelling results.

\end{document}